\documentclass[preprint,showpacs,preprintnumbers,amsmath,amssymb]{revtex4}

\usepackage{graphicx}
\usepackage{dcolumn}
\usepackage{bm}

\begin{document}


\title{FMR induced Josephson Current in a Superconductor/Ferromagnet/Superconductor Junction}

\author{S. Hikino$^{1}$}
\author{M. Mori$^{1}$}%
\author{S. Takahashi$^{1,2}$}
\author{S. Maekawa$^{1,2}$}
\affiliation{%
$^{1}$Institute for Materials Research, Tohoku University, Sendai 980-8577, Japan \\
$^{2}$CREST, Japan Science and Technology Agency, Kawaguchi 433-0012, Japan
}%



\begin{abstract}
We propose the phase dynamics induced by spin waves 
in a superconductor/ferromagnet/superconductor (SC/FM/SC) Josephson junction. 
The resistively shunted junction (RSJ) model, which describes the dynamics of superconducting phase difference, 
is extended to include the spin wave excitation by ferromagnetic resonance (FMR) 
using the gauge invariant phase difference between two $s$-wave superconductors. 
The current-voltage characteristics show step structures when the magnetization in FM is driven 
by tuning the microwave frequency to FMR in the SC/FM/SC junction. 
The result presents a new route to observe the spin wave excitation using the Josephson effect. 
\end{abstract}

\maketitle
The Josephson effect is characterized by a zero voltage current through a thin insulating barrier 
sandwiched by two superconductors\cite{josephson}. 
This effect is a macroscopic quantum phenomenon involving phase coherence between two superconductors. 
When a finite voltage-drop ($V$) appears in the junction, the phase difference ($\theta $) evolves 
according to $\partial \theta /\partial t = 2eV/\hbar $. 
As a result, an alternating current flows due to the time dependence of $\theta $. 
Moreover, when an alternating electric field due to microwave irradiation is applied to the junction, 
the current-voltage ($I$-$V$) characteristics show Shapiro steps\cite{shapiro}. 
Similar phenomena are observed in a normal metal weak link due to the proximity effect\cite{degennes, likharev}. 
The Josephson junctions are useful devices. 
For instance, 
the ac Josephson current is utilized to detect the electron-spin resonance in the Josephson junction doped with magnetic impurities\cite{barnes}.  
The dynamics of the Josephson junction which are reflected in the $I$-$V$ characteristics are described by the resistively shunted junction (RSJ) model, 
which is used to analyze Josephson dynamics\cite{stewart, maccumber}. 

The Josephson effect in superconductor/ferromagnet/superconductor (SC/FM/SC) junctions has been of 
considerable interest in recent years\cite{buzdin, golubov, ryazanov, kontos, oboznov, robinson, weides, mori}. 
One of the most interesting phenomena is the oscillation of the Josephson critical current ($I_{\rm c}$) 
in SC/FM/SC junctions. 
The origin of the oscillation is the exchange splitting of the conduction band in the FM, 
in which Cooper pairs penetrate into the FM with a finite center of mass momentum. 
If the thickness of the FM is about half of the period of the oscillation, 
the current-phase relation is shifted by $\pi $ from that of a conventional Josephson junction (0-junction). 
This is called a $\pi $-junction, which has a potential application as a quantum bit\cite{yamashita}. 
The $\pi$-junction is also used to measure experimentally a nonsinusoidal current-phase relation\cite{hsellier}. 
Most of the studies on SC/FM/SC junctions have so far been focused on the effect of the exchange splitting on 
Cooper pairs penetrating into the FM. 
Ferromagnetic materials possess dynamic properties such as spin waves, 
which may be excited by using ferromagnetic resonance (FMR). 
Therefore, the interaction between Cooper pairs and spin waves in the FM is expected to play 
an important role in the transport. 
Up to now, the spin dynamics in the FM has been disregarded in SC/FM/SC junctions. 

In this Letter, we study the effect of the spin dynamics on the Josephson effect in a SC/FM/SC junction. 
The RSJ model is extended to include the effect of spin wave excitations using the gauge invariant phase difference between 
superconducting leads. 
We adopt a model in which the magnetization of FM exhibits a precessional motion. 
The $I$-$V$ characteristic is calculated using the extended RSJ model. 
It is found that this characteristic exhibits step structures and 
the dc Josephson current is induced due to the spin wave excitation by FMR. 
The step structures appear whenever the relation $\omega_{\rm J} =2\ell \Omega $ is satisfied with $\ell $ being an integer, 
where $\Omega $ is the FMR frequency, and $\omega_{\rm J}=2eV/\hbar $. 
We propose a new route to observe spin waves using the Josephson current in SC/FM/SC junctions. 

\begin{figure}[b]
\begin{center}
\includegraphics[width=7.5cm]{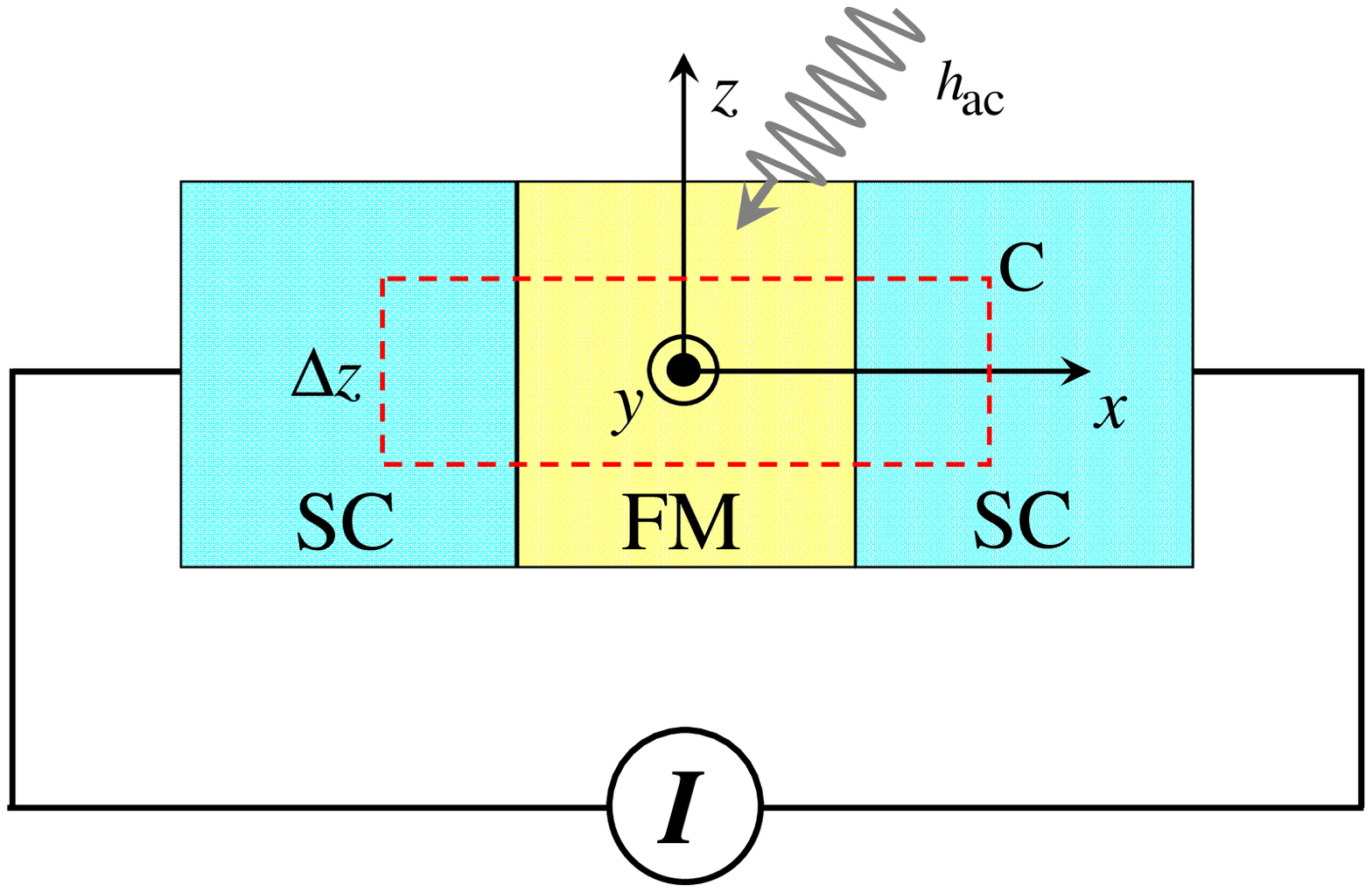}
\caption{Josephson junction with FM between two $s$-wave SCs. 
A microwave radiation $h_{\rm ac}$ onto FM causes the precession of the magnetization of FM. 
Coordinate system is chosen such that the electrodes surfaces are parallel to $yz$-plane. 
C is a contour in $xz$-plane.}
\label{setup}
\end{center}
\end{figure}

The system considered is a Josephson junction with a ferromagnet sandwiched by two $s$-wave superconductors. 
We choose a coordinate system such that 
the electrode surfaces are parallel to the $yz$ and $xz$ planes as shown in Fig.$^{~} $\ref{setup}, 
and the current flows along $x$. 
The Josephson junction is characterized by the phase difference $\theta (y,z,t)$ between SCs, 
which is described by the RSJ model as follows 
\begin{eqnarray}
i &=&
	i_{\rm c}^{0}\sin \theta \left(y,z,t \right)
	+\frac{1}{R}\frac{\Phi _{0}}{2\pi }
	\frac{d\theta \left(y,z,t \right)}{dt}
	+C\frac{\Phi _{0}}{2\pi }\frac{d^{2}\theta \left(y,z,t \right)}{dt^{2}},
\label{rsj}
\end{eqnarray}
where $i$ is an external current density and $i_{\rm c}^{0}$ is the critical current density. 
The flux quantum is denoted by $\Phi _{0}$. 
The resistance, $R$, and the capacitance, $C$, in the Josephson junction are normalized by the junction area, $S_{yz}$, 
as $R=R_{0}/S_{yz}$, $C=C_{0}/S_{yz}$. 
$R_{0}$ and $C_{0}$ are the resistance and the capacitance in the Josephson junction, respectively. 

\begin{figure}[t]
\begin{center}
\includegraphics[width=7.5cm]{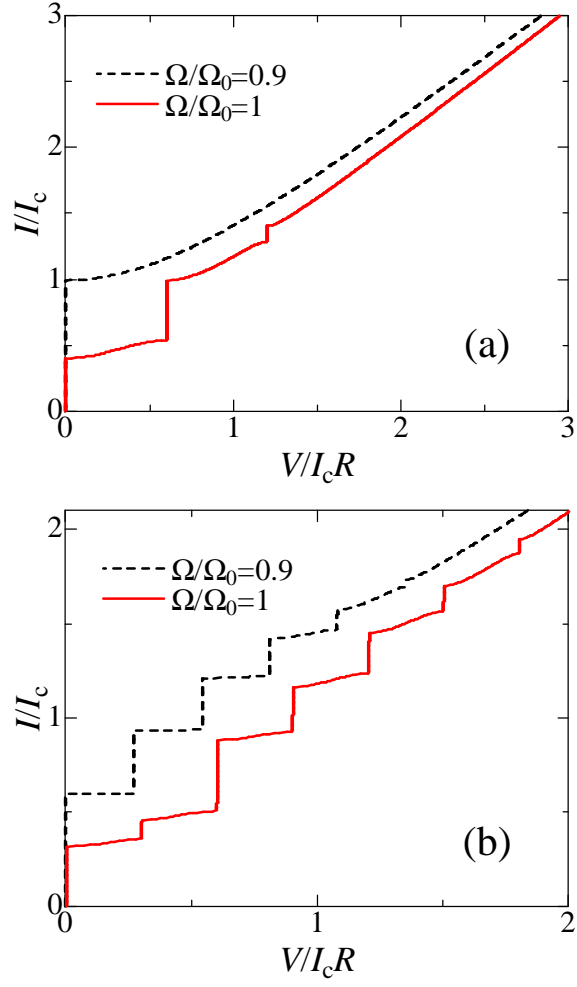}
\caption{Current-voltage characteristics. 
The solid and dashed lines show the $I$-$V$ characteristics microwave frequency 
for $\Omega /\Omega _{0}=0.9$ and $\Omega /\Omega _{0}=1$, respectively. 
We choose $\Phi _{\rm s}=0.03$. 
(a) $I_{\rm ac}/I_{\rm c}=0$. (b) $I_{\rm ac}/I_{\rm c}=0.6$. 
$I_{\rm ac}$ is ac current across the junction due to the electromagnetic field. }
\label{ivchara}
\end{center}
\end{figure}

We consider the situation, in which the FM is exposed to a circularly polarized microwave. 
The microwave causes the precessional motion of the magnetization, 
which corresponds to the excitation of the uniform mode of a spin wave. 
The motion of magnetization due to the microwave radiation is described by the Landau-Lifshitz-Gilbert (LLG) equation\cite{hillebrands}
\begin{eqnarray}
\frac{d{\bm M}}{dt} &=&
	-\gamma {\bm M}\times {\bm H_{\rm eff}}
	+\frac{\alpha }{M}
	\left[ 
	{\bm M}\times \frac{d{\bm M}}{dt}
	\right],
\label{mmeq}
\end{eqnarray}
where ${\bm M}\left(t \right)$ is the magnetization of the FM, $\gamma $ is the gyromagnetic ratio, and
$\alpha $ is the Gilbert damping. 
The effective field, to which ${\bm M}\left(t \right)$ responds, is given by 
${\bm H_{\rm eff}}={\bm H_{0}}+{\bm h_{\rm ac}\left(t \right)}$, 
where ${\bm H}_{0}$ is an uniaxial magnetic anisotropic field, which is parallel to $z$ axis and 
${\bm h_{\rm ac}}=\left(h_{\rm ac}\cos\Omega t,h_{\rm ac}\sin\Omega t,0 \right)$ is 
the microwave driving field with frequency $\Omega $. 
In Eq.$^{~} $(\ref{mmeq}), we neglect 
an anisotropic precession of the magnetization due to the demagnetization field for simplicity\cite{kittel, chikazumi}. 
The linearized solution of Eq.$^{~} $(\ref{mmeq}) is given by 
\begin{eqnarray}
M^{\pm }\left(t \right) &=&
	\frac{\gamma M_{z}h_{\rm ac}}{\Omega -\Omega _{0} \mp i\alpha \Omega }
	e^{\pm i\Omega t},
\label{mpm}
\end{eqnarray}
which describes the precession of $M^{\pm }\left(t \right)$ around ${\bm H_{0}}$ with frequency $\Omega$ 
and has a resonance with $\Omega_{0}=\gamma {H}_{0}$, 
where $M^{\pm}(t)=M_{x}(t)\pm iM_{y}(t)$. 
The rotating magnetic field is induced in the FM through the magnetic flux density ${\bm B}\left(t \right)=4\pi {\bm M}\left(t \right)$. 
Due to the magnetic field, the superconducting phase is not gauge-invariant. 
Therefore, we derive the equations for the gauge invariant phase difference between SCs. 
In the SC/FM/SC junction, the magnetic field in the FM is equal to $4\pi {\bm M}\left(t \right)$, 
since we assume that the two superconductors separated by the FM of the thickness $d$ 
are thick compared with the London's penetration depth 
and the magnetic field produced by the Josephson current is negligible. 
Therefore, the flux $\Delta \Phi_{y} $ parallel to the $y$ axis 
enclosed by a contour C in Fig.$^{~} $\ref{setup} is $\Delta \Phi_{y} = 4\pi M_{y} \left(t \right)d/\Phi _{0}\Delta z$. 
In a similar way, the flux $\Delta \Phi_{z}$ parallel to the $z$ axis is $\Delta \Phi_{z} = 4\pi Md/\Phi _{0}\Delta y$. 
Combining these relations with a gauge invariant phase difference, 
we obtain the differential equations with respect to $\theta \left(y,z,t \right)$\cite{tinkham, parks2}, 
\begin{eqnarray}
\nabla _{y,z}\theta \left(y,z,t \right) &=&
	-\frac{4\pi d}{\Phi _{0}}
	{\bm M}\left(t \right)\times {\bm n},
\label{dthetadt}
\end{eqnarray}
where $\bigtriangledown _{y,z} = \left(0,\partial/\partial  y,\partial/\partial  z \right)$, 
$\bm n$ is the unit vector perpendicular to the $yz$ plane. 
Equation $^{~} $(\ref{dthetadt}) is satisfied by the following solution, 
\begin{eqnarray}
\theta \left(y,z,t \right) &=&
	\theta \left(t \right)
	-\frac{4\pi M_{z}d}{\Phi _{0}}y
	+\frac{4\pi M_{y}\left(t \right)d}{\Phi _{0}}z, 
\label{thetayz}
\end{eqnarray}
where we adopt the gauge, such that the superconducting phase difference is equal to $\theta \left(t \right)$ without magnetic field. 
Introducing Eq.$^{~} $(\ref{thetayz}) in Eq.$^{~} $(\ref{rsj}) and integrating over the junction area, 
we obtain the extended RSJ model that includes the spin dynamics in the FM, 
\begin{eqnarray}
I/I_{\rm c} &=&
	\frac{\sin \left(\Phi _{\rm s}\tilde M_{y}\left(\tau \right) \right)}
	{\Phi _{\rm s}\tilde M_{y}\left(\tau \right)}
	\sin\left(\theta \left(\tau  \right) \right)
	+\frac{d\theta \left(\tau \right)}{d \tau }
	+\beta _{\rm c}\frac{d^{2}\theta \left(\tau \right)}{d \tau^{2} },
\label{exrsj}
\end{eqnarray}
with
\begin{eqnarray}
\Phi _{\rm s} &=&
	\frac{L_{z}d4\pi ^{2}}{\Phi _{0}}
	\frac{\gamma M_{z}h_{\rm ac}}{\Omega _{0}}, \\
\tilde M_{y}\left(t \right) &=&
	\frac{\left(\Omega/\Omega _{0} -1 \right)\sin\left(\Omega \tau _{\rm J} \tau \right)}
    {\left(\Omega/\Omega _{0} -1 \right)^{2}+\left(\alpha \Omega/\Omega_{0} \right)^{2}}
	+
	\frac{\alpha \Omega/\Omega_{0}\cos\left(\Omega \tau _{\rm J} \tau \right)}
    {\left(\Omega/\Omega _{0} -1 \right)^{2}+\left(\alpha \Omega/\Omega_{0} \right)^{2}}, 
\label{Phis&My}
\end{eqnarray}
where $I$ is an external current and 
$I_{\rm c}=I_{\rm c}^{0} \left( \pi \Phi _{0}/\Phi _{z} \right) \sin \left( \pi \Phi _{z}/\Phi _{0} \right)$. 
$I_{\rm c}^{0}$ is the Josephson critical current, and 
$\Phi _{z}$ is the total magnetic flux parallel to the $z$-axis in the FM. 
The other variables are given by, 
$\tau = t/\tau_{\rm J}$, $\tau_{\rm J}=\Phi _{0}/\left(2\pi I_{\rm c}R \right)$, 
and $\beta _{\rm c}=RC/\tau_{\rm J}$. When $\Phi _{z}=0$ and $\Phi _{\rm s}\tilde M_{y}\left(\tau \right)=0$, 
Eq.$^{~} $(\ref{exrsj}) reduces to a conventional RSJ model. 
The extended RSJ model is applicable to both 0- and $\pi $-junctions. 
Below, we solve numerically Eq.$^{~} $(\ref{exrsj}) for $\beta _{\rm c}=0$ (overdamped junction) 
and calculate the $I$-$V$ characteristic using the relation, 
$\left<\partial \theta/\partial t \right>=2e\left<V \right>/\hbar $, where $\left<\cdot \cdot \cdot \right>$ denotes a time average. 
For the numerical calculation, $\alpha $ is taken to be 0.01\cite{hillebrands}. 
First, we consider the situation that the magnetic field component of microwave is applied to the junction 
and the FMR occurs in the FM. 
This condition is realized by using a microwave cavity\cite{inoue}. 
The $I$-$V$ characteristic is calculated using Eq.$^{~} $(\ref{exrsj}) and 
the result is shown in the solid and dashed lines in Fig.$^{~} $\ref{ivchara}(a). 
The vertical axis is the normalized current, $I/I_{\rm c}$, and the horizontal axis is the normalized voltage, $V/I_{\rm c}R$. 
As shown by the dashed line in Fig.$^{~} $\ref{ivchara}(a), when the microwave frequency ($\Omega $) deviates 
from the ferromagnetic resonance frequency ($\Omega _{0}$), 
the $I$-$V$ characteristic is in agreement with that of a conventional Josephson junction. 
On the other hand, when the microwave frequency is in the condition $\mid1-\Omega /\Omega _{0} \mid \leq \alpha$ 
at which FMR occurs in the FM, 
the $I$-$V$ characteristic shows the step structure under the FMR as shown by the solid line in Fig.$^{~} $\ref{ivchara}(a). 

Second, we consider the case that the electromagnetic field due to microwave irradiation is applied to the junction. 
For the electric field component, 
we use the ac current source model in which $I$ is replaced by $I+I_{\rm ac}\sin \left(\Omega t \right)$ in Eq.$^{~}$(\ref{exrsj}), 
where $I_{\rm ac}$ is the amplitude of ac current\cite{baronetxt}. 
Figure$^{~}$\ref{ivchara}(b) shows the $I$-$V$ characteristic which reflects the electromagnetic field irradiation. 
In the dashed line in Fig.$^{~}$\ref{ivchara}(b), when $\Omega $ deviates from $\Omega _{0}$, 
the $I$-$V$ characteristic shows Shapiro steps due to the ac current similarly to conventional Josephson junctions. 
When $\Omega $ is equal to that of the alternating Josephson current, 
these steps arise at $V=n\frac{1}{2e}\hbar \Omega$ with $n$ being an integer. 
When the microwave frequency is in the condition $\mid1-\Omega /\Omega _{0} \mid < \alpha$, 
the step structure dramatically changes as
shown by the solid line in Fig.$^{~}$\ref{ivchara}(b). 
The amplitude of the second step is much larger than that of the first step. 
This behavior is different from that of the Shapiro step for which the amplitude of the steps decreases with increasing the voltage. 
In addition,  the step structures appear in the higher voltage region ($V/I_{\rm c}R>1.5$) unlike the case of the Shapiro steps. 

\begin{figure}[t]
\begin{center}
\includegraphics[width=7.5cm]{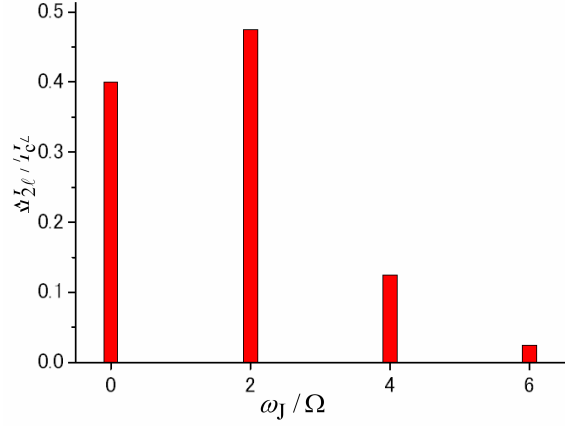}
\caption{The amplitude of the step in the $I$-$V$ characteristics vs the normalized Josephson frequency. 
The positions of step lie on the even number. 
$\Delta I_{2\ell }/I_{\rm c}$ is the amplitude of step in the $I$-$V$ characteristics, and $\ell $ is an integer. }
\label{ampofstep}
\end{center}
\end{figure}

To elucidate the origin of the step structure  in the $I$-$V$ characteristic due to the spin wave excitations, 
we analyze the first term of the Josephson current $I_{\rm J} \left(t \right)$ in Eq.$^{~} $(\ref{exrsj}). 
Using the generating function of Bessel functions and the standard mathematical expansion of the sine function, we have 

\begin{eqnarray}
I_{\rm J}\left(t \right) &=&
	-\frac{I_{\rm c}}{\zeta _{\rm s}}
	\sum_{n=0}^{\infty }
	\sum_{m=-\infty }^{\infty }
	J_{2m+1}\left(\zeta _{\rm s} \right)
	e^{i \left[ 
		\theta  \left(t \right)+2\left(n+m+1 \right)\Omega t +2\left(n+m+1 \right)\psi 
		\right]} \nonumber \\
		&&+\frac{I_{\rm c}}{\zeta _{\rm s}}
	\sum_{n=0}^{\infty }
	\sum_{m=-\infty }^{\infty }
	J_{2m+1}\left(\zeta _{\rm s} \right)
	e^{-i \left[ 
		\theta  \left(t \right)-2\left(n+m+1 \right)\Omega t+2\left(n+m+1 \right)\psi
		\right]},
\label{Jtstep}
\end{eqnarray}
where $J_{m}\left(\zeta _{\rm s} \right)$ is the Bessel function of the first kind, 
$\zeta _{\rm s}=\Phi _{\rm s}/\left(\Omega -\Omega _{0} \right)$, and 
$\psi =\arctan\left(\frac{\alpha \Omega}{\Omega -\Omega _{0}} \right)$. 
In Eq.$^{~} $(\ref{Jtstep}), when $\theta \left(t \right) = \pm 2\left(n+m+1 \right)\Omega t  $, 
the time averaged $I_{\rm J}\left(t \right)$ has a nonzero component of the dc Josephson current. 
As a result, the $I$-$V$ characteristics represent the step structure without changing voltage. 
Using the relation, $\partial \theta \left(t \right) /\partial t =\omega_{\rm J} $, we obtain the voltage at which the step structures occur,
\begin{eqnarray}
\omega_{\rm J} &=&
	2\left(n+m+1 \right) \Omega, 
\label{veqomega}
\end{eqnarray}
where $n$ and $m$ are integers. 
Figure$^{~}$\ref{ampofstep} shows the amplitude of the step structure 
as a function of the frequency $2\left(n+m+1 \right) \Omega$ in the $I$-$V$ characteristic. 
The vertical axis is the normalized amplitude of the step. 
It is seen that the step structures only appear at even number unlike the case of the Shapiro step. 

Finally, we discuss the condition $2(n+m+1) \Omega $ occuring step structures in the $I$-$V$ characteristic. 
We convert Eq.$^{~}$(\ref{veqomega}) into $V =2(n+m+1)\frac{1}{2e}\hbar \Omega $, 
which is quite similar to the condition $V=n\frac{1}{2e}\hbar \Omega$ of the Shapiro step, 
where the voltage is proportional to the number of photon $n$ and its energy $\hbar \Omega $\cite{shapiro, baronetxt}. 
In an analogous way, we interpret the above results as follows. 
In the present situation, a uniform mode of spin wave is excited in the FM by the FMR due to the microwave irradiation. 
Therefore, $ 2\left(n+m+1 \right)$ and $\hbar \Omega $ in Eq.$^{~} $(\ref{veqomega}) correspond to 
the number of quantized spin waves (magnons) and their energy, respectively. 
If Cooper pairs penetrating into the FM are scattered by the odd number of magnons, 
the Josephson coupling vanishes between the $s$-wave SCs because of the formation of a spin triplet state\cite{stakahashi}, 
resulting in no Josephson current. 
On the other hand, the Josephson coupling occurs 
when Cooper pairs penetrating into the FM compose the spin singlet state 
because of the even number of magnon scattering, so that the Josephson current flows. 
Therefore, the even number of magnons couples to the Josephson current. 
In conclusion, we have studied the effect of the spin dynamics on the ac Josephson effect 
in the superconductor/ferromagnet/superconductor (SC/FM/SC) junction. 
The resistively shunted junction (RSJ) model is extended to include the spin wave excitations induced by the ferromagnetic resonance 
(FMR) and the equations for the gauge invariant phase difference between SCs have been derived. 
We find that the current-voltage characteristics show step structures when the relation $\omega_{\rm J} =2\ell \Omega $ with 
$\ell $ being an integer is satisfied, where $\omega _{\rm J}=2eV/\hbar $ and $\Omega$ the FMR frequency. 
The results provide a new route to observe the spin waves using the Josephson current in SC/FM/SC junctions. 

The authors thank M. Aprili, I. Petkovic and S. E. Barnes for valuable discussions. 
This work is supported by a Grant in Aid for Scientific Research from MEXT, 
the next Generation Supercomputer Project of MEXT. 
The authors thank the Supercomputer Center, Institute for Solid State
Physics, University of Tokyo for the use of the facilities.


\newpage

\newpage

\newpage

\end{document}